\documentclass[12pt]{article}

\usepackage{latexsym}
\usepackage{amssymb}
\usepackage{amsfonts}

\usepackage{hyperref}
\usepackage[dvips]{graphicx,color}
\graphicspath{{images/}}

\def\nc#1{\newcommand{#1}}
\def\rnc#1{\renewcommand{#1}}

\def\a{\alpha}%GreekGreekGreek

\nc{\g}{\gamma}
\def\d{\delta}
\nc{\D}{\Delta} 
\nc{\e}{\eta}
\nc{\ep}{\epsilon}

\nc{\ve}{\varepsilon}
\nc{\G}{\Gamma}

\nc{\la}{\lambda}
\nc{\La}{\Lambda}
\nc{\om}{\omega}
\nc{\Om}{\Omega}
\nc{\vphi}{\varphi}
\nc{\si}{\sigma}
\nc{\Si}{\Sigma}
\rnc\th{\theta}
\nc\Th{\Theta}
\nc{\z}{\zeta}
\def\cO{{\cal O}}
\nc{\got}[1]{\mathfrak{#1}} %%%%%%%% gotic
\nc\im{{\rm Im}\, }
\nc\re{{\rm Re}\, }

\nc{\Rt}{{\tilde R}}
\nc{\CC}{{\mathbb C}}%%%%%%%%%%%% mathbb
\nc\II{{\mathbb I}} % math 1
\nc{\RR}{{\mathbb R}}
\nc{\HH}{{\mathbb H}}
\nc{\NN}{{\mathbb N}}
\nc{\ZZ}{{\mathbb Z}}
\nc{\MM}{{\mathbb M}}

\def\dag{\dagger}

\nc{\non}{\nonumber\\}
\def\nn{\nonumber}

\nc{\noi}{\noindent}

\nc{\p}{\partial}%Differentiations
\nc{\na}{\nabla}

\nc{\lan}{\langle}
\nc{\ran}{\rangle}

\nc{\beq}{\begin{equation}}
\nc{\eeq}{\end{equation}}
\nc{\beqa}{\begin{eqnarray}}
\nc{\eeqa}{\end{eqnarray}}
\nc{\beqas}{\begin{eqnarray*}}
\nc{\eeqas}{\end{eqnarray*}}
\nc{\barr}{\begin{array}}
\nc{\earr}{\end{array}}
\nc{\ben}{\begin{enumerate}}
\nc{\een}{\end{enumerate}}
\nc{\bit}{\begin{itemize}}
\nc{\eit}{\end{itemize}}

\nc\xb{{x^\dag}}
\nc\xt{{\widetilde{x}}}
\nc\xtb{{\widetilde{x}^\dag}}
\nc{\Xt}{{\widetilde{X}}}
\nc{\Xtb}{{\Xt^\dag}}
\nc\Lat{{\widetilde {\La}}}
\nc\tb{{\overline t}}
\nc\xdx{{\Xb X}} 
\nc\Mb{{\bar M}} 
\nc\Tb{{\bar T}}
\nc\vev[1]{\lan #1\ran}
\nc\refeq[1]{(\ref{#1})}
\nc\rQ{\mbox{ \textcolor{red}{$\leftarrow$(?)}}}
\nc\more{{ \textcolor{red}{MORE}}}

\rnc\tb{{\overline t}}

\def\cO{\mathcal{O}}

\def\ZZ{\mathbb{Z}}
\def\RR{\mathbb{R}}
\def\CC{\mathbb{C}}

\def\zdef{\mkern-1mu\,\hbox{$\,\raise.3pt\hbox{:}\mkern-5mu =\mkern1mu\,$}}

\def\pd{\partial}

\def\re{\mathrm{Re}}
\def\im{\mathrm{Im}}

\newcommand{\eqref}[1]{(\ref{#1})}%numerowanie równañ w nawiasach

\def\nn{\nonumber}

\def\half{\frac{1}{2}}

\def\Lat{\widetilde{\La}}
\def\Tb{\overline{T}}

\def\Xb{X^\dag}

\def\Xt{\widetilde{X}}

\def\xt{\widetilde{x}}

\begin{document}
\title{Non-minimal Gauge Mediation and Moduli Stabilization} 

\author{T. Jeli\'{n}ski,  Z.  Lalak and J. Pawe{\l}czyk\\{\small {}}\\{\small {\it Institute of Theoretical Physics, University of Warsaw,}}\\{\small {\it ul.\ Ho\.za 69, 00-681 Warsaw, Poland}}\\}

\date{}
\maketitle

\abstract{
\addtolength{\textwidth}{-2cm}\small In this paper we consider $U(1)_A$-gauged Polonyi model with two spurions coupled to a twisted closed string modulus. 
This offers a consistent setup for metastable SUSY breakdown
 which allows for moduli stabilization and naturally leads to gauge or hybrid gauge/gravitational mediation mechanism. 
Due to the presence of the second spurion one can arrange for a solution of the $\mu$ and $B_\mu$ problems in a version of modified Giudice-Masiero mechanism, which works both in the limit of pure gauge mediation and in the mixed regime of hybrid mediation.}

%\hoffset -2.5cm
%\textwidth 18cm
%\addtolength{\textwidth}{3.5cm}
\section{Introduction}
Gravity mediation and gauge mediation are two well understood benchmark schemes for the 
transmission of supersymmetry
breaking from the hidden to the visible sector. Each of these  schemes has its
virtues. Gravity is a universal messenger, present in all spontaneously broken locally supersymmetric models. The goldstino is higgsed away and 
a relatively straightforward cancellation of the cosmological constant may take place. Perhaps
the most difficult question in gravity mediation is the suppression of 
flavour changing neutral currents, while
in the gauge mediation schemes   it is difficult to get the $\mu$ and $B_\mu$ 
parameters  in the range appropriate
for the electroweak symmetry breaking. Dominant gauge mediation offers a light gravitino with interesting cosmology and observable low energy signatures. In general one expects a mixture of both schemes to be at work in realistic models. Therefore, it is of great interest to assess theoretical possibilities leading to consistent realizations of supersymmetry breaking schemes leading to sizable contributions of the gauge mediation mechanism. 
Complete models for gravity/gauge mediation have recently been analyzed  \cite{sspot,Nomura,Lalak:2008bc,Dudas}. In \cite{Lalak:2008bc} a variety of O'Raifeartaigh models coupled to a simple messenger sector and to gravity has been studied. The spurion has been assumed to be a gauge singlet, and a general conclusion is that in these models the dominance of gauge mediation is rather natural. On the other hand, in \cite{Dudas}  the spurion has been charged under an anomalous 
$U_A (1)$ symmetry. The gauge variation of the this superfield is compensated in the superpotential by a gauged shift of an untwisted K\"ahler modulus. In this case, for a change, it has been found that the gauge mediation dominance is rather hard to achieve. This result is somewhat disappointing in the context of stringy models, where the anomalous $U_A (1)$ symmetries are a common phenomenon. 
  
Hence, it is important to find out a consistent string-motivated  framework where pure gauge mediation or hybrid gauge mediation can be naturally realized. 
A promising setup is given by D-brane constructions 
\cite{lotsof}. In particular 
in \cite{CW}  stabilization and supersymmetry breakdown in a gauged O'R setup is partially due to a correction to the superpotential and partially to anomalous D-terms.

Here we consider two Polonyi type spurion  superfields and investigate a complete model with modulus and D-term contributions to the scalar potential included. Contrary to \cite{Dudas} our closed string modulus is the twisted one \cite{quivers} which  has dramatic impact on the dynamics. Also, it appears that corrections to the K\"ahler potential \cite{Kitano} are crucial  for stabilization of  both spurions. In passing to the effective low energy  model the gauge field and the modulus become heavy and decouple. The effective low energy superpotential  has an approximate global $U(1)_A$ symmetry broken explicitly by the terms linear in spurion superfields, no Fayet-Iliopoulos term gets generated \cite{Komargodski:2009pc}.

 In a sense, the setup with two independent spurions is very general and the example we consider gives a rather complete description of phenomenological options provided by a wide class of models with metastable supersymmetry breakdown. In particular, we consider a possibility of generating effective operators which can produce $\mu$ and $B_\mu$ terms even in 
the regime of dominant gauge mediation. In the region of hybrid gauge/gravitational mediation the standard
 Giudice-Masiero terms may be forbidden by gauge invariance, since the spurions are no longer gauge singlets. However, we have found a set of effective operators replacing the standard ones, which work satisfactorily supplying the correct size $\mu/B_\mu$ terms even in the domain of hybrid mediation.

\section{Single spurion model}

In this section we describe a simple single spurion model realizing the gauge mediatation scenario. This superfield will be charged under a $U_A(1)$.
The model under consideration is in fact a slight modification of the standard Polonyi models. It is well known that the linear term of the superpotential in order to be $U_A(1)$ invariant must couple to a closed string modulus. The crucial point is that, contrary to \cite{Dudas}, the modulus we are going to consider is the so-called twisted K\"ahler modulus. 

\subsection{Effective Lagrangian}
\label{oneX}

We start with the simple Polonyi model supplemented by a correction to the
K\"ahler potential \cite{Kitano}:
\beq\label{Pol}
K=|X|^2-\frac{|X|^4}{\La^2}, \quad W=W_0+ f X .
\eeq  
The quartic piece of (\ref{Pol}) can have various origins which shall be discussed at the end of this  section e.g. it may arise  from  an exchange of 
chiral multiplets with a supersymmetric mass term proportional to $\La < M_P$, called thereafter  rafertons.
In order to cancel the cosmological constant one  needs to choose $f=\sqrt{3} W_0$. With this choice the model (\ref{Pol}) has a non-SUSY minimum at 
\beq\label{X-P}
X=\frac{\La^2}{2\sqrt3}+\mathcal{O}(\La^4).
\eeq

One easily finds (see the next section) that for $\vev X \ll 1$ the gauging of this model does not change  the above conclusion
substantially. Let 
\beq\label{Pol-g}
K=|X|^2-\frac{|X|^4}{\La^2}+\frac{\la^2}{2} (T+T^\dag- V)^2, \quad W=W_0+f X e^{-T},
\quad D=|X|^2-2\frac{|X|^4}{\La^2} + \lambda^2 (T + T^\dag).
\eeq 
The scale of the mass parameter $\la$ which sets the mass of the gauge field is model dependent and can range from the string scale $M_s$ down to values a few orders of magnitude below $M_s$ (\cite{anommass}). 
The non-SUSY minimum is located at 
\beq\label{g-X-P}
X=\frac{\La^2}{2\sqrt3\left(1+\frac{3}{2}l^2\right)}, \quad T =-\frac{l^2\Lambda^2}{24\left(1+\frac{3}{2}l^2\right)^2},
\eeq
where $l=\frac{\La}{\la}$. The masses of %$(\textrm{Re}(X),\textrm{Im}(X),\textrm{Re}(T))$ 
excitations around vevs are
\beq
\left(\frac{2\sqrt{2}f}{\La}\sqrt{1+\frac{3}{2}l^2},\,\frac{2\sqrt{2}f}{\La}\sqrt{1+\frac{3}{2}l^2},\,2\la\right).
\eeq
%\[
%
%\]
The residual value of the $D$ term at the above minimum is  $\sim \frac{f^2}{\la^2} $. This can give a mass of the order of $\frac{m_{3/2}}{\la}\sim \frac{f}{\la} $ to charged scalars in the visible sector, but since we are interested in mediation schemes where gravity participates at the level of $10 \%$ or less, such a contribution remains subdominant (see Section \ref{phenom}). 

It is interesting to note, that this way one could also 
stabilize, unfortunately - under somewhat restrictive  
conditions - a pair of charged untwisted moduli, 
say $M_1$ and $M_2$. Let's assume for simplicity that 
both moduli transform with the same shift under $U(1)_A$, 
i.e. have the same charge. 
Then we can form a combination, say ${ S}$, of these 
superfields which is neutral, and the second one, $T$,
 which has the charge twice 
as large as that of $M_i$. Then the D-term will contain a 
dynamical FI term of the form 
$-\partial K / \partial {T} ({ S},{T})$ for 
a generic K\"ahler potential. At this point the stabilization 
of the neutral field ${S}$ can be achieved by a racetrack 
mechanism, and the stabilization of the charged field 
$T$ can be done as above, with the help of the D-term. 
To obtain the correct size of $f$ we assume that it depends on an expectation value of an  uncharged modulus stabilized in a standard way by means of a racetrack mechanism. 

The correction to $K$ can be thought of as a remnant  after integrating out heavy rafertons, see for example \cite{Lalak:2008bc}.
In the UV-complete model one starts with a model of the form 
\beq
W=W_0+f X e^{-T}+\frac{1}{2} \Phi_1^2 X+m \Phi_1\Phi_2,
\eeq
\beq
K= \frac{\la^2}{2} \left ( T + T^\dag -V \right )^2 + \Phi_{1}^{\dagger} e^{\frac{1}{2} \,  V} \Phi_1 + \Phi_{2}^{\dagger} e^{-\frac{1}{2} \, V} \Phi_2  + X^\dagger e^{- V} X \, .
\eeq 
Integrating out rafertons $\Phi_1,\Phi_2$ down to some scale $Q^2$ yields a correction to the K\"ahler potential (we take the scale $Q$ in such a way, that 
the K\"ahler potential for $X$ is canonical)
\beq\label{kitano}
\d K= -\left(\frac{1}{192 \pi^2 m^2}  + \frac{1}{\la^2} \right)
 (X^\dag X)^2. 
\eeq
The second term comes from the tree-level (dominant) contribution of the anomalous gauge boson and can be of  importance depending on a value of $\la$. Moreover this contribution is generic and may be the source of stabilization of $X$ in the absence of candidate rafertons. 
As for the rafertons, their expectation values stay zero in the presence of the complete 1-loop correction.
Since the resulting expectation value for the twisted modulus is very small, negligible with respect to $\langle X \rangle$, and both the modulus and the $U_A(1)$ gauge boson are heavy, the phenomenology of the resulting low-energy model is the same as that of the models discussed by \cite{Lalak:2008bc}.

In the above we have implicitly focused on twisted moduli, whose vacuum expectation value could be small. The story of untwisted moduli, which suffer from the run-away problem, is more complex, and does not lead to satisfactory results, as explained in \cite{Dudas}.

\section{Non-minimal model} 

In what follows we want to examine a string inspired supergravity model of gauge mediated (metastable) supersymmetry breaking with dynamically stabilized K\"ahler modulus. The hidden sector of this  model consists of an anomalous $U(1)_A$ gauge group, two chiral matter $X$, $\Xt$ with opposite (unit) charges under $U(1)_A$ and twisted K\"ahler modulus $T$. All these ingredients are related to intersecting D-branes configurations. Dynamics of that model is controlled by superpotential $W$ 
\beq
W=W_0+f Xe^{-T} + \widetilde{f} \widetilde{X} e^{+T}-\frac{1}{M}(X\Xt)^2
\eeq
which has a non-perturbative "anomalous" part originating from D-instantons
\cite{Witten96,lotsof}
 and a non-renormalizable  $(X\Xt)^2$ part. The latter is an indispensable term originating from  the tree level exchange of heavy string states. Thus we expect that $M$ is of the order of the string scale:  
$M\sim M_s\leq1$ .  
As we shall see this new contribution provides a nontrivial  
modification  of the dynamics of the  moduli stabilization.
The inverse sign in front of T (compared to the standard $e^{-S_{inst}}$) appearing in $\widetilde{f} \widetilde{X} e^{+T}$ is forced by the opposite charge of $\Xt$ and it may arise as
an instanton effect coming from a D-brane wrapping a cycle with modulus  $S-T$ where $S$ is an uncharged modulus. Then $e^{-S}$ gets a vev and enters our definition of $\widetilde f$.
The other constants
$W_0,\ f,\ \widetilde{f}$
may also depend on some  moduli which we assume to be stabilized and are heavy enough not to influence the physics discussed here.
 We also tune the cosmological constant to vanish by taking  ${\rm max} \{f,  \widetilde{f} \} \sim W_0$.
 
The K\"ahler potential $K$ is a direct generalization of \refeq{Pol-g} with corrections \refeq{kitano}.
It reads
\begin{eqnarray}
K=X^\dag X+\Xt^\dag \Xt+\frac{\la^2}{2}(T+T^\dag)^2 +\d K_R + \d K_A,
%K=Se^{-V_a+V_b}S^\dag+\St e^{V_a-V_b}\St^\dag +qe^{V_a}q^\dag+\qt e^{-V_b}\qt^\dag+\half(T+\Tb-\d (V_a-V_b))^2
\end{eqnarray}
where
\beq \label{kr}
\d K_R=-\frac{1}{\La^2}(X^\dag X)^2-\frac{1}{\Lat^2}(\Xt^\dag \Xt)^2
\eeq
results from integration over rafertons (see \refeq{kitano})
and 
\beq \label{ka}
\d K_A=-\frac{1}{\la^2}(X^\dag X-\Xt^\dag \Xt)^2
\eeq
comes  from integration over the 
anomalous $U(1)_A$ gauge field.
We have added subscripts to $\d K$ to make distinction between the two situations.
The scenarios associated with the two corrections  differ in details because of the K\"ahler mixing involved in $\d K_A$. 
 One should keep in mind  that $W$ and $K$ of this form describe a consistent model in the regime $X$, $\Xt\ll\La, \ \Lat$ and $X\Xt\ll{M}^2$. 

Our main result is that if we take the above model, including twisted $T$ modulus and using 1-loop corrections to K\"ahler potential,  then in fact we get metastable SUSY breaking minimum with fixed $\vev X,\ \vev T$ which satisfies phenomenological constraints. The crucial requirement is that both coefficients of the quartic terms in (\ref{kr}) are negative. Thus e.g. the model of \cite{CW} does not respect this requirement  as there 
is only one instanton term, i.e. $\widetilde f=0$ there.
The usage of the twisted moduli here 
contrasts with the usage of untwisted moduli in \cite{Dudas}, 
where small $\vev X$ implies large $\vev T$ which in turn kills the $F_X$ term and makes the realistic supersymmetry breaking difficult.

There is also a part of the scalar potential  due to the D-term
\[
V_D=\half D^2=\half \left(|X|^2-|\Xt|^2+\la^2(T+T^\dag)\right)^2,
\]
where we have suppressed $\cO(X^4,\Xt^4)$ terms and the dependence on the coupling constants of the anomalous $U(1)_A$ gauge group.  One should notice that the presence of $T$ relaxes the condition $|X|=|\Xt|$ (from $D=0$) used e.g. in \cite{CW}.
%rescaling

Let us make some general statements about  the role $T$ plays in the dynamics. While $\la$ can be close to the string scale we take $W_0$ to be much smaller, of the order of $f$, to be in accord with phenomenology. This immediately leads to the conclusion that if $\la\sim1$ then $D=0$ is a very good approximation for the equation of motion (EOM) for 
 $T$. It is also apparent that $T$ is very massive: $m^{2}_T\approx \la^2$.  In addition  $\vev T\ll 1$ so in the first approximation we can suppress $T$ in the  EOMs for both $X$'s. Thus the procedure  to  find  the vacuum of the system is to find solution for $\vev X$ disregarding the D-term and then to determine the  $\vev T$ from $D=0$. 
The effective low energy superpotential has an approximate global $U(1)_A$ symmetry broken explicitly by the terms linear in spurion superfields, no Fayet-Iliopoulos term gets generated \cite{Komargodski:2009pc}. At low energies the modulus $T$ is stabilized and frozen. On the other hand if $\la\sim\La$ then one has to take into account F-term part of e.o.m for $T$, i.e. now $\left<T\right>$ can be found from $\pd_TV_F+\la^2 D=0$, where $V_F$ is F-term part of the potential.
 
\newpage
\section{Metastable SUSY breaking minima}

In this section we shall 
derive formulae for the minuma of the SUGRA potential under consideration. We shall consider two cases: $\vev X\sim \vev \Xt$ and $\vev X\gg \vev \Xt$ 
which can be treated analytically. 
As it shall be discussed at  the end of the paper the solutions give different physics. The crucial ingredient that makes it possible to  create 
phenomenologically realistic minima in the scalar potential is the correction to the K\"ahler potential. The two types of 
$\d K_{R,A}$ originating form the exchange of rafertons and the "anomalous" $U(1)_A$ gauge boson appear to have  different impact on dynamics.

In what follows we shall use the rescaled scalar potential
$v=Ve^{-K} /f^2 \approx V/f^2$ and rescaled variables $x=X/\a,\ \xt=\Xt/\a$, $w_0=W_0/f,\ \a=\sqrt[3]{\frac{f M}{2}}, \ep=\widetilde{f}/f$.
 Notice that $w_0\sim 1$. In this notation one obtains 
$F_x/f =1-x\xt^2 + m_{3/2} x^\dag \alpha$, $F_{\xt}/f =\epsilon-x^2\xt + m_{3/2} \widetilde{x}^\dag \alpha$, where $m_{3/2}=e^{K/2} \langle W \rangle \sim  \langle W \rangle$.

\subsection{$x\approx \xt$} \label{reg1}

Here we go to the regime where fields $x,\xt$ are so small that 
the dominant terms in the scalar potential  are similar to those for the single $X$ model discussed in Sec. \ref{oneX}. Taking $\d K=\d K_R+\d K_A$ one obtains\footnote{The $w_0=\sqrt{\frac{1+\ep^2}{3}}$ and we set $\Lat=\La$ for simplicity.}:
%\beq
%v=\alpha \left (-2 \epsilon  \overline {\tilde {x}} - 2 \bar {x} - 
%   2 \epsilon  \tilde {x} - 2 x \right) + 
% \frac {\a^2 |x|^2} {w_ 0^2 \La^2}+\frac{\a^2 \ep^2 |\xt|^2 }{w_ 0^2 \Lat^2}
%\eeq
\beqa
v&=&-2w_0\a\left(x+x^\dag+\xt\epsilon+\xt^\dag\epsilon\right)+\frac{\a^2}{\La^2}|x|^2\left[4+(6-3\epsilon^2)l^2\right]+\frac{\a^2}{\La^2}|\xt|^2\left[4\epsilon^2+(6\epsilon^2-3)l^2\right]\nonumber\\
&&-3\epsilon\frac{\a^2}{\la^2}\left(x^\dag\xt+x\xt^\dag\right),
\eeqa
where $l=\frac{\La}{\la}$. The minimum of $v$ is located at
\beq\label{min}
X=\La^2\,\frac{4\ep^2+3(3\ep^2-1)l^2}{h},\quad\Xt=\ep\La^2\,\frac{4+3(3-\ep^2)l^2}{h}.
\eeq
$h$ is defined as follows:
\[
h=\frac{24\ep^2-16(\ep^4-4\ep^2+1)l^2-27(\ep^2-1)^2l^4}{\sqrt{3}\sqrt{1+\ep^2}}.
\]
Here $F_X\sim F_{\Xt}\sim f$ and masses of the excitations around vevs are:\footnote{Two states have a mass  $m_-$ each and two have $m_+$ each. The mass of the twisted modulus is $m_T=2\la$.}
\[
m_{\mp}=\frac{f}{\La}\left(4 \left(\epsilon^2+1\right)+3\left(\epsilon^2+1\right)l^2\mp\sqrt{16 \left(\epsilon^2-1\right)^2+56l^2
   (\epsilon^2-1)^2+\left(49\epsilon^4-62\epsilon^2+49\right)l^4}\right)^{\frac{1}{2}}.
\]
The $m_-^2$ is positive iff
\[
\frac{\La}{\la}<\frac{2\ep}{\sqrt{3}|\ep^2-1|}.
\]
As one can see there are no minima with $\la<\frac{\sqrt{3}|\ep^2-1|}{2\ep}\La$, the reason being the extra terms in $v$ originating from  the mixing between $X$ and $\Xt$ in $\d K_A$.  

The validity of the approximation requires 
\begin{equation} \label{consist1}
\La\ll\a^{3/4}\sim (f M)^{1/4}.
\end{equation}
For instance, with $f \sim 10^{-12},\ M\sim 10^{-1} $ this  gives $\La < 10^{-13/4}$. These values of parameters correspond to the (upper) border of the region of gauge mediation dominance.

\subsection{$\xt\sim 1/{x^2}$}  \label{reg2}

Here we are looking for  extrema in the region
of the run-away rigid supersymmetry  vacuum i.e. 
$x^2  {\widetilde x}\sim 1$, while $x\gg 1$. 
In this case there is not much difference between $\d K_R$ and 
$\d K_A$ because the mixing terms are of the order $x \xt$ and hence  subleading. In general,  contributions to the K\"ahler metric that are proportional to $\widetilde{x}$ are negligible in this case. 
\beqa
v&=&|\epsilon-x^2 \xt|^2-2w_0(x +x^\dag)\alpha  + 
\left(\frac {4}{\La^2}+\frac{6-3\ep^2}{\la^2}\right)\alpha ^2 |x|^2
\eeqa
Notice that $\Lat$ does not enter the formula for $v$.
Critical point respects:
\beqa
0&=& 2 x {x^2}^\dag\xt\xt^\dag-2\ep x\xt -2w_0\a+ \a^2\left(\frac {4}{\La^2}+\frac{6-3\ep^2}{\la^2}\right)x^\dag\nn\\
0&=& x^2{x^\dag}^2\xt^\dag - x^2 \epsilon
\eeqa
The second equation yields ${\widetilde {x}}=\frac{\ep}{x^2}$ while the first one gives $x\approx\frac{w_0\La^2}{2\a\left(1+\frac{6-3\ep^2}{4}l^2\right)}$. Thus the physical fields get vevs
\beq
X=\frac{\La^2}{2\sqrt{3}\left(1+\frac{6-3\ep^2}{4}l^2\right)},\quad \Xt=\frac{6\ep fM\left(1+\frac{6-3\ep^2}{4}l^2\right)^2}{\La^4}.
\eeq
Masses of the excitations are: 
\[
\left(\frac{2\sqrt{2}f}{\La}\sqrt{1+\frac{6-3\ep^2}{4}l^2},\,\frac{\La^4}{3\sqrt{2}}\frac{1}{\sqrt{1+\frac{6-3\ep^2}{4}l^2}},\,2\la\right).
\] 
%(\frac {f^2 \epsilon ^2} {w_0^4 \Lambda ^4} + \frac{64 w_ 0^2 \ \Lambda ^8}{M^2})^{1/2}$.
One checks that the validity of the approximation requires 
\begin{equation} \label{consist2} 
\la\gg\a^{3/8}\sim (f M)^{1/8},
\end{equation} \label{smallf}
 i.e. this minimum exists for rather big values of $\la$. One finds
 \begin{equation}\label{Fs}
F_X=f,\quad F_{\widetilde{X}} = m_{3/2} \widetilde{X}\sim\frac{f^2M\ep}{\La^4} \ll f .
\end{equation}
Of course there exists also another minimum with appropriate 
replacement of $X$ and $\Xt$.

As we have seen the reliability of our analytic calculations puts 
severe constrains on 
the parameter space of the model. The interesting question is what happens beyond the literally taken region of validity of our analytical approximation. 
The answer  goes beyond the scope of this paper, however numerical study confirms that useful solutions do exist. 

In the following section devoted to  the phenomenological analysis we shall not treat the obtained numerical limits very strictly, rather we shall consider  them as useful guiding posts.  

%\newpage
\section{Phenomenological implications}
\label{phenom}

\subsection{Hierarchy of visible scales}
Presence of a second superfield participating in the supersymmetry breaking raises questions about the hierarchy of various mass scales transmitted to the visible sector. First of all, let us notice, that the gravitino mass is always determined by the largest F-term, which we take to be $F_X$
\begin{equation}
m_{3/2} \sim  \frac{max(F_X, F_{\widetilde X})}{\sqrt{3} } 
=\frac{F_X}{\sqrt{3} }\approx \frac{f}{\sqrt{3} }\, .
\end{equation}
However, the  gauge mediation mechanism can be  sourced by either  $X$ or $\widetilde{X}$. Also, the question of the generation of $\mu$ and $B_\mu$ can have a more complicated answer. Let us discuss these two problems one after another. 

The sourcing of the gauge mediation occurs via the coupling of the supersymmetry breaking spurion superfield to messengers 
$q, \, Q$:
\begin{equation}
\delta W = \lambda_1 X q Q + \lambda_2 \widetilde{X} q Q. 
\end{equation}
Obviously, because of the $U(1)_A$ gauge invariance, only one of the above couplings can be nonvanishing. 
Once we source the gauge mediation and gravity mediation by the single superfield, the story develops in the standard fashion, as discussed in 
\cite{Lalak:2008bc}. The important remark is that using the suitable value of the larger of the $f$-parameters and the suitable 
%$\d K$ 
scale $\Lambda$, 
one can smoothly interpolate between hybrid gauge-gravity mediation when   $\Lambda^2$ is of the order of $10^{-3} \, M_P$, and gauge domination for smaller values of that scale. 

Lets assume for a while that it is the smaller of the two F-terms which sets the gauge mediation, say $F_{\widetilde{X}}$, which means $\lambda_1 =0 $. 
The thing to notice is that there is a second universal gravity scale which can set nearly universal soft terms, the scale of anomaly mediation: 
\beq
m_{(a)} = \frac{\alpha}{4 \pi} m_{3/2}.
\eeq  
This can happen for scalars which have no-scale type K\"ahler potential or for gauginos without the tree-level interactions with the Polonyi superfield. 
Hence there is a limited space for the atypical situation with gravitino and - hence - uncharged moduli somewhat heavier than the gauge mediated soft masses:
\begin{equation} \label{eq:27}
\frac{\alpha}{4 \pi} m_{3/2} < m_{sf} < m_{3/2} \, .
\end{equation}
To realize this situation microscopically we need 
$\frac{\alpha}{4 \pi} F_X  < \frac{\alpha}{4 \pi} F_{\widetilde{X}} / \widetilde{X} < F_X$. 
In the first class of solutions, see section \ref{reg1}, 
this  results in the condition 
$ \widetilde{\Lambda} > 10^{-3/2} \sqrt{{\widetilde f}/f} $. 
The requirement of the $1$ TeV gaugino masses, 
$m_{sf} = \frac{\alpha}{4 \pi} F_{\widetilde{X}} / \widetilde{X} 
\approx 1\, {\rm  TeV}$,  implies via (\ref{eq:27}) 
$10^{-15}<f<10^{-12}$. The auxiliary condition 
(\ref{consist1}) leads via the above to the relation 
$\widetilde{\Lambda}  <  10^{-13/4}$.
All these relations become compatible when 
${\widetilde{f}}/{f} < 10^{-7/4}$,
hence  the non-standard hierarchy, $m_{sf} < m_{3/2}$ can be
 realized. On the other hand, sourcing the gauge mediation by
  $F_X$ through (\ref{eq:27}) is impossible. 
Analogous result holds for the second solution, section
 \ref{reg2}.   

Despite the non-standard option, from the point of view of required GM dominance, and naturalness,  it is rather  obvious that it is the largest F-term which should actually source the gauge mediation of supersymmetry breakdown to the visible sector. Hence,  the following we shall assume that $F_X>F_{\Xt}$ hence we shall set $\lambda_1 \neq 0$, $\lambda_2 =0$.

\subsection{Solution to  $\mu$ and $B_\mu$ problems}

The easiest way out of the $\mu$ and $B_\mu$ problems would be to rise the scale of the gravity mediated contribution to the soft masses to the level of about 10\% and to call for the Giudice-Masiero mechanism \cite{GiuMa}. 
The basic Giudice-Masiero mechanism of generating the $\mu$-term in the effective low-energy superpotential relies on the presence in the high-energy 
K\"ahler potential of an interaction term of the form $\delta K_\mu  = \frac{1}{2} X^\dagger  H_u H_d \, + \, {\rm h.c.}$.
With the help of the well known formulae, see \cite{Kaplunovsky:1993rd}, one obtains this way 
\begin{equation}
\mu = \left | m_{3/2} X^\dagger - F^{X^\dag} \right | = m_{3/2} \left  | X^\dagger - \sqrt{3} \right | \, .
\end{equation}
The additional term\footnote{Note that in fact there is the Planck scale suppression in these operators: $\delta K_\mu  = \frac{1}{2} \frac{X^\dagger}{M_P} H_u H_d \, + \, {\rm h.c.}$ and $\delta K_B = \frac{1}{2} \frac{X X^\dagger}{M^{2}_P} H_u H_d \, + \, {\rm h.c.} $.}$\delta K_B = \frac{1}{2} X X^\dagger H_u H_d \, + \, {\rm h.c.} $ gives $B_\mu$: $B_\mu = \pm  m_{3/2} \, \mu$.
Hence, if one choses the gravitino mass to be of the order of $100$ GeV, one finds $\mu$ and $B_\mu / \mu$ also of this order of magnitude, which, as shown in  
\cite{Lalak:2008bc}, may support the radiative electroweak breaking mechanism. 

However, this basic proposal works well if the spurion $X$ is a gauge singlet. If the spurion is charged, the two operators $\delta K_\mu , \, \delta K_B$ 
cannot be allowed simultaneously. Also, since there is some physics at scales $\Lambda$, $\widetilde{\Lambda}$, one expects that operators suppressed by a scale smaller than $M_P$ are created. Since the properties of the vacuum of our model are known, we can check whether the presence of two spurions does allow for new operators solving the $\mu / B_\mu$ problem.

Since we are assuming that there are heavy rafertons, or heavy gauge bosons, with the characteristic mass scale $\Lambda$ or $\widetilde{\Lambda}$, we 
can write down the following effective operators, assuming correct $U(1)_A$ charges, 
\begin{equation} \label{mubmu}
{\cal{O}_{\mu}} = \eta  \left ( \frac{{{X}}^{\dagger} H_u H_d}{{\Lambda}} \right )_D, \, {\cal{O}_{B}} = 
\left ( \Xt H_u H_d \right )_F \, .
\end{equation}
Note, that since the charges of $X$ and $\widetilde{X}$ are opposite, only one of the operators $\widetilde{X}^{\dagger} H_u H_d$ and $X^{\dagger} H_u H_d$
is allowed, and the same is true for the operator ${\cal{O}_{B}}$. The coupling $\eta$ takes into account the fact, that the required correction to 
the K\"ahler potential may be borne via loop diagrams, hence the additional suppression of this operator is very likely to take place. Hence, the coupling 
$\eta$  could be as small as $10^{-1} - 10^{-3}$. 

The operators (\ref{mubmu}) produces the $\mu$-term and  the $B_\mu$:
\begin{equation}  \label{roles}
{\cal{O}_{\mu}} \rightarrow \eta\frac{F^*_{{X}}}{{\Lambda}} \left ( H_u H_d \right )_F,\quad
{\cal{O}_{B}} \rightarrow F_{\Xt} \left ( H_u H_d \right )_A \, .
\end{equation}
Correct electroweak breaking prefers{}\footnote{This condition is of course a very crude estimate, and a detailed analysis may reveal other interesting solutions. In particular, note that we have not used the $\ll$ operator.}: $B_\mu \leq \mu^2$, hence 
\begin{equation} \label{eq:136}
F_{\Xt} \leq  \left(\frac{F_X}{{\Lambda}/\eta }\right)^2.
\end{equation}
Now let us request that the $\mu$-term is of the order $100$ GeV, which means in Planckian units $\eta\  F_X/{\Lambda} = 10^{-16}$ 
i.e.
%$F_X=f=10^{-16}{\La}/{\eta}$ and from \refeq{eq:36} 
we get $F_{\Xt}\leq 10^{-32}$. 
In the case $x\approx \xt$ this can be reached  with 
extremely small $\widetilde{f}\approx 10^{-32}$. Requesting that the $F_X$ gives the gauge mediated soft masses of $1$ TeV leads to the condition
$\Lambda = 10^{-4} / \eta$, which is somewhat outside the range of the original analytical approximation. However, such a solution may turn out in numerical studies. Another possibility to find a solution in this region is to exchange\footnote{This means changing the charge of the operator $H_u H_d$. } the roles of $X$ and $\Xt$ in (\ref{roles}) (while keeping $F_X \geq F_{\Xt}$). In such a case the  condition (\ref{eq:136}) gives 
 $\widetilde{\Lambda} \leq \eta \sqrt{\widetilde{F}}$. This prefers unrealistically small scales $\widetilde{\Lambda}$.

On the other hand in the case $x\gg \xt$ (with $ w_0\approx 1$) from Eq. \eqref{Fs}:
\beq
\frac{ f^2 M \ep}{\La^4}\leq\left(\frac{f}{{\Lambda}/\eta }\right)^2\ \to
\La^2\geq\frac{ M\ep}{\eta^2}.
\eeq
Taking into account that we expect a small $\eta$, we also need small $\ep$ in order to achieve  reasonable $\La$. For instance,  with  
$M=10^{-2}$, the $\eta=10^{-1}$ requires $\ep\leq\La^2$.

Numerical checks going beyond the analytic approximations show, that the realistic vacuum solution does exist in this case.  
For instance, numerical study confirms the existence of the following solution
\begin{equation}
\La=10^{-2},\ M=10^{-1},\ f=10^{-14},\ w_0=1=\ep \,
\end{equation}
which gives 
\begin{equation}
x=20,\ \xt=2\cdot10^{-3},\
F_x=f,\ F_{\xt}=6\cdot 10^{-5} f \ .
\end{equation}

\section*{Summary and conclusions}

We have  investigated non-minimal gauge invariant models for F-term supersymmetry
breaking in gauge mediation and hybrid gauge/gravity mediation scenarios.
In Section 2 the gauge invariant stabilisation of the Polonyi-type spurion field in the presence of the charged untwisted modulus has been described in a simple model. 
Both the modulus and the $U(1)_A$ gauge boson become heavy and practically decouple from the low-energy degrees of freedom, leaving a model with the explicitly broken global $U(1)$ symmetry and giving rise to the phenomenology described in \cite{Lalak:2008bc}. We have demonstrated  that also a pair of untwisted moduli can be stabilized in this framework. 
This analysis extends and complements  the discussion of the role of untwisted moduli in gauge invariant models given in \cite{Dudas}, pointing out how the problems raised there can be efficiently circumvented. 
Sections 3.-4. extend the discussion of the gauge invariant stabilization and supersymmetry breakdown to models with a pair of spurion superfields.
It has been demonstrated that corrections to the K\"ahler potential are crucial for the stabilization of the spurions and thus for triggering the supersymmetry breaking.  
The phenomenological features of such extended models have been discussed in section 5. It has been concluded, that also in the presence of charged moduli in the supersymmetry breaking sector one can easily arrange for gauge mediation dominance and also for a hybrid gauge/gravity mediation, which naturally avoids the supersymmetric flavour problem. We have shown that one can use the presence of two spurion fields to arrange for operators rather naturally supplying the $\mu$ and $B_\mu$ terms of the correct size to support acceptable electroweak breaking. 
Our setup can be realized in string theoretical constructions and in F-theory models. 

%\newpage
\section*{Acknowledgments}
\vspace*{.5cm}
\noindent Authors thank Michele Trapletti and Alberto Romagnoni for illuminating and  helpful discussions. We would like to acknowledge  stimulating discussions with Emilian Dudas, Stefan Pokorski and Krzysztof Turzy\'nski.  
\noindent This work was partially supported by the 
EC 6th Framework Programme MRTN-CT-2006-035863 and  by TOK Project  
MTKD-CT-2005-029466. JP and  ZL  thank CERN Theory Division  for hospitality.

\end{document}